\documentclass[12pt]{article}

\usepackage{amsfonts}
 \usepackage{amssymb}
 \usepackage{amsmath}

\usepackage{graphicx}

\def\pmx{\begin{pmatrix}}
\def\emx{\end{pmatrix}}
\def\bsq{\begin{subequations}}
\def\esq{\end{subequations}}

\newtheorem{thm}{Theorem}
\newtheorem{cor}[thm]{Corollary}
\newtheorem{lemma}[thm]{Lemma}

        \def\pf{\medbreak\noindent{\bf Proof:}\enspace}
     \def\rmk{\medbreak\noindent{\bf Remark:}\enspace}

\def\be{\begin{eqnarray}}
\def\ee{\end{eqnarray}}
\def\bee{\begin{eqnarray*}}
\def\eee{\end{eqnarray*}}
\def\ds{\displaystyle}

\def\bra{\langle}
\def\ket{\rangle}

\def\bm{{\bf m}}
\def\det{{\rm det}}
\def\av{{\rm av}}

\def\supp{{\rm supp}}
\def\mm{ \! - \!}
\def\wtd{\widetilde}
\def\up{\Upsilon}

\def\nn{\nonumber}

\def\raw{\rightarrow}

\def\1rt2{{\textstyle \frac{1}{\sqrt{2}}}}
\def\'{^{\prime}}

\title{One-dimensional models for atoms \\in  
strong magnetic fields, II:
\\ Anti-Symmetry in the Landau Levels 
 \thanks{\copyright by authors.}}

\author{Raymond Brummelhuis \\
School of Economics, Mathematics and Statistics  \\
Birkbeck College
- University of London \\
Mallet Street, London, United Kingdom
  \\ {\normalsize r.brummelhuis@statistics.bbk.ac.uk }
\and Mary Beth Ruskai\thanks{The work of MBR was partially supported  by
  the National Science
        Foundation under Grants DMS-0074566 and DMS-0314228.} 
   \\ Department of Mathematics , 
Tufts University\\
  Medford, Massachusetts 02155 \\ {\normalsize marybeth.ruskai@tufts.edu}}

\date{\today \\ ~~ \\ Dedicated to Elliott Lieb on the occasion of his
70th birthday}

\begin{document}

\maketitle

\begin{abstract}
Electrons in strong magnetic fields can be described
by one-dimensional models in which the Coulomb potential
and interactions are replaced by regularizations associated
with the lowest Landau band.   For a large class of models
of these type, we show that the maximum number of electrons
that can be bound is less than $a Z + Z f(Z)$.   
The function $f(Z)$
represents a small non-linear growth which reduces to 
$A_p Z (\log Z)^2$ when the
magnetic field $B = O(Z^p)$ grows polynomially with 
the nuclear charge $Z$.   
In contrast to earlier work, the models considered here
include those arising from realistic cases in which
the full  trial wave function  for N-electrons  
is the product of an $N$-electron trial function
in one-dimension and an antisymmetric product
of states in the lowest Landau level.

\end{abstract}



\section{Introduction}

It is well-known that systems in strong magnetic fields behave 
like systems in one-dimension, i.e., a strong magnetic field  
confines the particles to Landau orbits orthogonal to the
field, leaving only their behavior in the direction of the field
subject to significant influence by a static potential.
Motivated by this general principle and the work of Lieb,
Solovej and Yngvason \cite{LSY} (LSY) on atoms in extremely 
strong magnetic fields, Brummelhuis and Ruskai \cite{BR1} 
initiated a study of models of atoms in homogeneous strong magnetic
fields in which the 3-dimensional wave-function  has the form
\begin{equation} \label{eq:psi.lowland}
  \Psi({\bf r}_1, {\bf r}_2 \ldots {\bf r}_n) = 
      \Phi(x_1 \ldots x_n) \, \Upsilon(y_1,z_1, y_2,z_2, \ldots y_n,z_n)
\end{equation}
where $\Upsilon$ lies in the projection onto the lowest Landau
band for an N-electron system.  We follow the somewhat non-standard
convention of choosing the magnetic field in the x-direction,
i.e, ${\bf B} = (B, 0, 0)$ where $B$ is a constant denoting
the fields strength, in order to avoid notational confusion
with the nuclear charge $Z$.

The Hamiltonian for an $N$ electron atom in a magnetic field ${\bf B}$ is 
\begin{equation} \label{eq:Ham.3dim}
H(N,Z,B) = \sum_{j=1}^N \left[ |{\bf P}_j + {\bf A}|^2  
     -\frac{Z}{|{\bf r}_j|} \right] +
     \sum_{j<k} \frac{1}{|{\bf r}_j - {\bf r}_k |} 
\end{equation}
where  ${\bf A}$ is a vector potential such that   
  ${\bf \nabla}  \times  {\bf A} = {\bf B}$. The ground-state energy of   
$H(N, Z, B ) $ is given by   
\begin{equation}  \label{eq:E0}
E_0(N,Z,B) = \inf_{\| \Psi \| = 1} 
  \langle \Psi , H(N, Z, B )  \Psi \rangle 
\end{equation}
Let $E_0^{\rm conf}(N,Z,B)$ denote the corresponding infimum
restricted to linear combinations of functions of the form
(\ref{eq:psi.lowland}). 
For extremely strong fields, it was shown in \cite{LSY} 
 that $E_0 / E_0^{\rm conf} \rightarrow 1$ as 
$B/Z^{4/3} \rightarrow \infty$ with $N/Z$ fixed.

In this paper  we consider $E_0^{\Upsilon}(N,Z,B)$,  the infimum 
when (\ref{eq:E0}) is further restricted to those functions
of the form (\ref{eq:psi.lowland}) corresponding  to a 
particular choice for $\Upsilon$.
As discussed in \cite{BR1}, is straightforward to show that
\begin{equation}  \label{eq:scale}
  E_0^{\Upsilon}(N,Z,B) = \sqrt{B} \inf_{\| \Phi \| = 1} 
    \langle \Phi , h(N, Z, B^{-1/2} )  \Phi \rangle  + N B
\end{equation}
where
\begin{equation}\label{eq:1dimHam}
h^{\Upsilon}(N,Z,M) = 
\sum_{j=1}^N \Big[
   - \frac{1}{M} \frac{d^2}{dx_j^2} -Z \widetilde{V}_j^{\Upsilon}(x_j)
\Big] +  \sum_{j< k} \widetilde{W}_{jk}^{\Upsilon}(x_j- x_k),
\end{equation}
and we have scaled out the field strength $B$ so that
the only remnant of the magnetic field is in the ``mass''  
$M = B^{-1/2}$.  The effective one-dimensional potentials
$\widetilde{V}_j^{\Upsilon}$ and $\widetilde{W}_{jk}^{\Upsilon}$  
can be written in terms of the functions \cite{BR1,BRW,SS}
\be  \label{eq:Vm.def}
V_m(x) = \frac {1}{m!} \int_0^\infty 
        \frac{s^{2m} \, e^{-s^2}}{\sqrt{x^2 + s^2}} \, s \, ds ,
\ee  
which are discussed in Section~\ref{sect:Vm}  
and studied in detail in \cite{RW}.
The precise form of  $\widetilde{V}_j^{\Upsilon}$ and
$\widetilde{W}_{jk}^{\Upsilon}$ depends on the choice of
$\Upsilon$; some special cases are discussed in
Section~\ref{sect:mag.model}.   When
 $\Upsilon$ is a simple product of one-particle Landau
states, or a finite linear combination of such
products, the effective
potentials  satisfy $\widetilde{V}_j^{\Upsilon}(x) \leq V_{\mu}(x)$ and
$\widetilde{W}_{jk}^{\Upsilon}(x) \geq V_{\nu}(x) $ for some
integers $\mu$ and $\nu$ which depend upon $\Upsilon$.  
We will primarily be interested in the case of symmetrized and
anti-symmetrized products of one-particle Landua states.
In this case, as discussed in more detail in Section~\ref{sect:Vm}  
and Appendix~\ref{apx}, bounds of the form above are readily
obtained.   For those situations in which
 $\nu < 2\mu$, a bound on the maximum negative ionization is given
in Theorems~\ref{thm:main} and \ref{thm:log}.  

In \cite{BR1} we considered the simple, but unrealistic
situation in which $\Upsilon$ is a product of 
Landau states with $m = 0$.  In this paper we introduce a
more realistic model, which we call the ``Slater model'', in which
$\Upsilon$ is an antisymmetrized product of Landau states.
As in \cite{BR1}, we concentrate on the question of maximum
negative ionization.   Define $N_{\max}(Z,B)$ as 
 the maximum number of
electrons for which the Hamiltonian (\ref{eq:1dimHam}) has
a bound state in the sense $E_0(N,Z,B) < E_0(N-1,Z,B)$.

LSY \cite{LSY} showed
that in extremely strong magnetic fields, atoms  bind $2Z$ electrons in
the sense
\begin{equation} \label{eq:max2}
  \liminf_{Z, \, B/Z^3 \rightarrow \infty} \frac{N_{\max}(Z,B)}{Z} \geq 2.
\end{equation}
and conjectured that $2Z$ was also an upper bound to this limit.
However, even for the simple model in \cite{BR1} 
we were only able to show  the  weaker bound
$N_{\max}(Z,B) < 2Z + 1 + c \sqrt{B}$.
Unfortunately, when $B > O(Z^3)$ as required
for the limit in (\ref{eq:max2}), the term
$c \sqrt{B} = c Z^{3/2}$ dominates so that
we can only conclude that 
$  N_{\max}(Z,B)  \leq 2Z + O(Z^{3/2})$.

In this note we show that for a large class of
one-dimensional  models,
including  some in which $\up$ in (\ref{eq:psi.lowland})
is a simple product
 or a Slater determinant, the bound on $N_{\max}(Z,B)$
can be improved to one of the form $aZ + Z f(Z)$  
with $f(Z) = O(\log Z)^2$ when $B$ grows polynomially
with $Z$.    Our results were announced earlier in \cite{BR2,BRW} 
in the form 
\be  \label{eq:gform}
N_{\max}(Z,B) \leq aZ + Z g(Z,B)
\ee
 with $g(Z,B) = (\log Z)^2 + \log Z (\log B)^{1 + \omega}$ for some
$\omega > 0$.  Subsequently, Seiringer
\cite{S} gave a similar bound for the full 3-dimensional
Hamiltonian. 
For fermions, Seiringer's bound \cite{S} has  the form
(\ref{eq:gform}) with
$g(Z,B) = \min \big\{ \big( \frac{B}{Z^3} \big)^{2/5}, 
\big( \log \frac{B}{Z^3} \big)^2 \big\}$; it is 
obtained by applying Lieb's method to the full 3-dimensional
Hamiltonian.  Hainzl and Seiringer \cite{HS}
then extended this bound  to a density
matrix model in which the variable perpendicular to the field is
replaced by discrete angular momentum quantum numbers.


Our use of one-dimensional models was motivated by a
desire to understand the physics associated with the
consequence of the one-dimensional character of
atoms in strong magnetic fields, which is well-known
and made precise in the work of LSY \cite{LSY}. 
 Brummelhuis and Duclos
\cite{BD1,BD2,BD3} also showed that, for each fixed total
angular momentum in the field direction, the full QM Hamiltonian
(\ref{eq:Ham.3dim}) converges in norm-resolvent  sense\footnote{To be
precise, let $R $ be  the resolvent of $H $, and
$R_s $ that of $H_s $.  Then   
$   \| \Pi_s R \Pi_s - R_s \| = O(B^{-1/2})$ on $\Pi_s {\cal H}$,  
and $    \| R  \| = O(B^{-1/2})$ on $\Pi_s^{\perp} {\cal H}$ at a
distance $(\log B)^2$ to the spectrum of $H_s $.} 
to the projected Hamiltonian 
$H_s (N, Z, B ) = \Pi_s H(N, Z, B ) \Pi_s $ where
 $\Pi _s $ denotes the orthogonal projection onto
the lowest Landau band. 
In the special case of  zero total angular momentum,
$\bra \Psi H_s (N, Z, B )  \Psi \ket$ has the form  (\ref{eq:1dimHam})
when $\Psi$ has the form (\ref{eq:psi.lowland}). Full details will be
given in \cite{BD3}.   In contrast to \cite{LSY}, the strategy in
\cite{BD1,BD2} does not require an a-priori bound on
$N $ in terms of $Z $, but {\em does} need to fix the total angular
momentum in the direction of the magnetic field.

Despite Seiringer's result \cite{S} and the work in \cite{HS}, 
we feel that our argument, which uses the RS  localization approach, 
is of some interest.  Because
our analyses of the models in \cite{BR1} showed that electrons
are highly delocalized in the direction orthogonal to the field,
it may seems surprising that such a localization technique
works at all.  However, a careful examination of the proof
in section 3, shows that it reflects this delocalization in
the sense that the ``inner ball''  grows with $B$.  
Nevertheless, the localization error can be controlled
with a modest   excess charge as described in the results which
follow.

We now summarize our results in the theorems which follow.
\begin{thm}  \label{thm:main}
Suppose the potentials in the Hamiltonian (\ref{eq:1dimHam})
satisfy 
\be   \label{eq:eff.bd}
\widetilde{V}_j^{\Upsilon}(x) \leq   V_{\mu}(x)  \quad \hbox{and} \quad
 \widetilde{W}_{jk}^{\Upsilon}(x) \geq  \1rt2 V_{\nu-1}
\Big(\frac{x}{\sqrt{2}} \Big), 
\ee 
for all $j,k$ and  $0 \leq   \nu \leq 2 \mu$.
  Then  
for every $\alpha > 0$  
 there is a constant $A_{\alpha} > 0$ and  constants $a_1, a_2$
(independent of $\alpha$) 
such that the Hamiltonian $h^{\Upsilon}(N,Z,B)$ has no bound states provided that
\bsq	\be
  \label{eq:wklowbd}    N > 
 2   Z +   A_{\alpha } Z^{1 + \alpha} , \quad \hbox{and}
\\  
 a_1 e^{Z^{\alpha/4 }}    > B  \geq a_2 Z^{\gamma_{\nu}} ,
\ee 
\esq
where the exponent $\alpha$ can be arbitrarily small and
 the exponent $\gamma_{\nu}$  depends upon $\nu$.  
In particular, when $ \nu = O(1)$, it suffices to take
$\gamma_{\nu} > 2$; when $\nu = O(N)$ one must choose
   $\gamma_{\nu}  > 3$.
\end{thm}
Although the non-linear term is higher order than
$2 Z$, it is useful to write the linear
term separately.  It is due to the relative strength of 
the potentials near the nucleus, while the non-linear terms
are  needed to control the localization error.
The upper bound  on $B$ is needed for technical reasons
associated with the fact that the localization error can
not be controlled when $B$ grows exponentially. 
As discussed in Remark 1 of Section~\ref{sect:finpf},
the requirement $\nu \leq 2 \mu$ can be relaxed at the
expense of replacing $2Z$ by $cZ$ in (\ref{eq:wklowbd})  with $c > 2$.


The non-linear term in the lower bound (\ref{eq:wklowbd}) can  be
improved to one that is logarithmic.  We first state it 
in general and then under  
  the simple, and realistic, assumption that $B = Z^p$
for $p > 3$.  
The case $\gamma_{\nu} > 3 $ in  (\ref{eq:loglowbd}),
corresponds to the superstrong region considered 
by LSY in \cite{LSY}.
\begin{thm}  \label{thm:log}
Assume that the potentials 
$\widetilde{V}_j^{\Upsilon}$ and
$\widetilde{W}_{jk}^{\Upsilon}$ satisfy (\ref{eq:eff.bd})
with  $0 \leq   \nu \leq 2 \mu$.
 Then there are positive
constants $A, a_\epsilon, a_2$ such that
the Hamiltonian $h^{\Upsilon}(N,Z,B)$ has no bound states provided that
\bsq	\be
  \label{eq:wklowbd2}    N > 
 3   Z +  1 + A \, Z \log Z \,  |\log \tfrac{Z^2}{B} | ,
\quad
\hbox{and}
\\  \label{eq:loglowbd}
 a_{\epsilon} e^{Z^{1/2} - \epsilon}  > B  \geq a_2 Z^{\gamma_{\nu}} ,
\ee  \esq
where $\epsilon > 0$ can be arbitrarily small and $\gamma_{\nu}$
is as in Theorem~\ref{thm:main}.
\end{thm}
\begin{cor}  \label{thm:Zp}
Assume that the potentials 
$\widetilde{V}_j^{\Upsilon}$ and
$\widetilde{W}_{jk}^{\Upsilon}$ satisfy (\ref{eq:eff.bd})
with  $0 \leq   \nu \leq 2 \mu$
and that $B = aZ^p$ for some  $a > 0$ and
$p > 3$.
  Then there is a
constant $A$ such that
the Hamiltonian $h^{\Upsilon}(N,Z,B)$ has no bound states provided that
  \be   \label{eq:Zpbd}
  N >   3   Z + A Z (\log Z)^2.
\ee   
\end{cor}

Each of these results, immediately yields an upper bound on
$ N_{\max}(Z,B) $ which we state for ease of comparison with
the results in \cite{BRW,HS,S}.   Under the hypotheses of
Theorem~\ref{thm:log}
\be
 N_{\max}(Z,B) \leq 3Z + A \, \log Z \, |\log \tfrac{Z^2}{B} |
\ee
Under the hypotheses of Corollary~\ref{thm:Zp}
\be
 N_{\max}(Z,B) \leq 3Z + A Z (\log Z)^2.
\ee
Unfortunately, unlike Theorem~\ref{thm:main}, the $3Z$
in the linear term includes a contribution from the
localization error as well as the expected $2Z$ from
electrostatics.


\section{Effective potentials} \label{sect:mag.model}

\subsection{Regularized Coulomb potentials}  \label{sect:Vm}

The Landau state with energy $B$ and angular momentum $-m $
can be written compactly using the complex variable
$\zeta  = y + iz$ as
\be  \label{eq:landau}
\gamma_m^B(y, z ) =   
[\pi m!]^{-1/2}B^{(m+1)/2} \overline{\zeta}^m e^{-B |\zeta |^2 /2}.    
\ee
The effective one-dimensional potentials in our models
can be written using the regularization of the 3-dimensional
Coulomb potential with a Landau state, i.e.,
\be
  V_m^B(x) & \equiv & \langle \gamma_m^B , \frac{1}{|{\bf r} |} \, 
        \gamma_m^B \rangle  =   
     \int_{{\bf R}^2} \frac{|\gamma_m^B |^2}{|{\bf r}|} dy \, dz \nonumber
\\
     & = & \frac {B^{m+1}}{m!} \int_0^\infty 
        \frac{s^{2m} e^{-Bs^2}}{\sqrt{x^2 + s^2}} \, s \,ds  \nonumber
\\
     & = &  \frac{1}{ m!} \int_0^\infty 
          \frac{u^m e^{-u}}{\sqrt{x^2 + u/B}} du  \label{vmdef} \\
    & = & \frac {2B^{m+1}}{m!} e^{Bx^2}\int_{|x|}^\infty  
             (t^2 - x^2)^m e^{-Bt^2} dt  ,  \nonumber
\ee
where ${\bf r}$ in ${\bf R}^3$ and $s = y^2 + z^2$.
In view of the rescaling in (\ref{eq:scale}), it suffices
to consider only the case $B = 1$ for which we drop the
superscript, i.e, $V_m(x) \equiv V_m^{1}(x)$.
The properties of $V_m(x)$ were studied in detail
in \cite{RW} and summarized in \cite{BRW}.  Those which we need here
are listed below.
\be  \label{eq:Vm.prop.a}
V_m(x) ~\hbox{is monotonically decreasing for} ~  x \geq 0.
\ee
\be  \label{eq:Vm.prop.b}
V_{m + 1}  (x) < V_m  (x) < \dfrac{1}{|x|}.
\ee
\be  \label{eq:Vm.prop.c}
\dfrac{1}{\sqrt{x^2 + m}} >   V_m  (x)  >
 \dfrac{1}{\sqrt{x^2 + m  + 1}} 
\ee
\be \label{eq:Vm.prop.d}   \hbox{If} ~   
 V_{\av} (x) \equiv \frac{1}{N} \sum_{j = 0}^{N - 1} V_j  (x) ,
~\hbox{then} ~  V_{\av} (x) \leq 2V_N  (x).
\ee

 \subsection{Simple product Landau model} 
We restrict to wave   
functions of the form   (\ref{eq:psi.lowland}) with
$\Upsilon = \prod_{k=1}^N  \gamma_{m_k}^B(y_k,z_k) $
a simple product of Landau states. 
Then   
 \be\ \label{1}   
\Psi_{m_1 \ldots m_N} = 
      \Phi(x_1 \ldots x_n) \prod_{k=1}^N  \gamma_{m_k}^B(y_k,z_k) ,  
\ee
and   
\be 
 \langle \Psi_{m_1 \ldots m_N}, H(N,Z,B ) \Psi_{m_1 \ldots m_N} 
    \rangle  =  \sqrt{B} \,
 \langle \Phi, \, h^{\bm}(N,Z,B^{-1/2}) \Phi \rangle  + NB      
\ee  
where we rescale as in (\ref{eq:scale}) and 
  $ {\bf m} = (m_1, \cdots , m_n ) $.  Then
\begin{equation}  \label{eq:effHam}
  h^{\bm} (N,Z,M ) =   
    \sum_{j=1}^N \left[ - \frac{1}{M}\frac{d^2}{dx_j^2} -Z V_{m_j}(x_j)
\right] +   
         \sum_{j< k} W_{m_j,m_k}(|x_j - x_k|)  , 
\end{equation} 
 $V_m $ is
given by (\ref{eq:Vm.def}) (with $B = 1$) and  the
effective interaction satisfies 
\be   \label{eq:Weff}
W_{m, m^{\prime}} (x - x' ) & =  &  
\langle \gamma_m \otimes \gamma_{m'} ,   
  { \frac{1}{|{\bf r} - {\bf r'} |}} \, 
     \gamma_m \otimes \gamma_{m'} \rangle   
 \\
    & = &  
   \sum_{j=0}^{m+ m^{\prime}} b_j \frac{1}{\sqrt{2}}   V_j
      \Big( \frac{|x - x'|}{\sqrt{2}} \Big)  \label{eq:conv.mm} \\
   & \geq & \frac{1}{\sqrt{2}}  V_{m+ m^{\prime}}   
    \Big( \frac{|x - x' |}{\sqrt{2}} \Big) 
  \label{eq:W.ineq}
\ee
for some $b_j \geq 0$  with $\sum_j b_j = 1$.  
That $W_{m, m^{\prime}} $ can be written as a convex sum as in
(\ref{eq:conv.mm}) was shown in \cite{PRWRH};  For completeness, 
a proof   in the special case $m = m\'$
is included in the Appendix.
 When   $m = m' = 0$  (\ref{eq:conv.mm}) reduces to 
$W_{0, 0} (|x - x' |) =   
\frac{1}{\sqrt{2}}  V_0\Big( \frac{|x - x' |}{\sqrt{2}} \Big) $  
as shown in \cite{BR1}.   

When all $m_j = m $ are equal, we denote the
effective Hamiltonian in (\ref{eq:effHam}) by
$h^m(N,Z,M)$ and refer to it as the
{\em $m$-momentum Landau model}.   For this model,
$\wtd{V}^{\up}(x) = V_m(x)$ and 
$\wtd{W}^{\up}(x) = \1rt2 V_{2m} \Big(\frac{x}{\sqrt{2}} \Big)$.
The case $m = 0$ was considered in \cite{BR1}
and \cite{BRW}.

\begin{cor} \label{cor:0}  Let $h^m(N,Z,B^{-1/2}) $ be the
Hamiltonian described above.  Then for any
$\alpha > 0 $ and any $B $ satisfying   
$a_1 e^{Z^{\alpha/4 }}  > B  \geq a_2 Z^{2+\varepsilon} $ for 
suitable constants $a_1$ $a_2$ and some $\varepsilon > 0$, there   
exists a constant $A_{\alpha} > 0 $    
\begin{equation}   
N_{\max} (Z, B ) \leq   2Z + A_{\alpha} Z^{1 + \alpha} .
\end{equation}   
Moreover, when $B = a Z^p$ for some $p > 3$,  one can find a constant
$A$ (depending on $a$, $p$) such that
\begin{equation}   
N_{\max} (Z, B ) \leq   3Z + A Z ( \log Z)^2 .
\end{equation}   
\end{cor}

\medskip

The Thomas-Fermi theories introduced by LSY in \cite{LSY} for the
superstrong and hyperstrong regimes have a    kinetic energy term
typically associated to bosonic systems. It seems therefore reasonable   
to consider
$h^m(N,Z,B^{-1/2} )$ with domain in the symmetric wave functions.  This
is however    in clear contradiction with the fact that the electrons
described by the original $3$-dimensional
Hamiltonian (\ref{eq:Ham.3dim}) are fermions, and
(\ref{eq:psi.lowland}) should be anti-symmetric.    Therefore,
in the next section, we introduce a model which reflects the
anti-symmetry.

\subsection{A Slater determinant Landau model}   \label{sect:slat}   

It is reasonable to expect   that the    electrons will try to satisfy the
Pauli principle by going into different orbits in the lowest Landau 
band, and that any realistic one dimensional model will have similar
behavior. We now consider the special case in which  $\Psi$ has the form
(\ref{eq:psi.lowland}) with $\up$ an anti-symmetrized product 
constructed using $m = 0, 1, 2 \ldots N \mm 1$.  Thus, we let
\begin{equation}   
\Upsilon =   
\left( \frac{1}{\sqrt{N!}} \gamma_0 \wedge \cdots \wedge \gamma_{N - 1} \right),   
\end{equation}  
where the wedge $\wedge$ denotes the anti-symmetric product 
so that $\up$ is  a Slater determinant in  the Landau states 
$\gamma_j$ for   $j = 0 , 1 \ldots N \! - \! 1$.   
In this case,   
\begin{equation}   
\langle H(N, Z, B ) \Psi , \Psi \rangle =   
\sqrt{B} \, \langle h^{\det}(N,Z,B^{-1/2})   \Phi , \Phi \rangle  + NB  
\end{equation}   
with   
\begin{equation}
   h^{\det}(N,Z,M) =
    \sum_{j=1}^N \left[ - \frac{1}{M}\frac{d^2}{dx_j^2} -Z V_{\av} (x_j)
   \right] +  \sum_{j< k}   
         W_{\det} (|x_j - x_k|) ,   
\end{equation}   
where   
\begin{equation}  \label{eq:V.Slat} 
V_{\av} (x) = \frac{1}{N} \sum_{j = 0}^{N - 1} V_j (x)   
\end{equation}   
and   the effective interaction  is 
\be  \label{eq:Wdet}
 W_{\det} (x) = \1rt2
     \sum_{j = 0}^{N - 2} b_{2j + 1} 
   V_{2j + 1}\Big( \tfrac{x}{\sqrt{2}} \Big)
\ee
with    $ b_{2j + 1} \geq 0 $ and 
$\ds{\sum_{j = 0}^{N - 2} b_{2j + 1} = 1}$.
It then follows from  (\ref{eq:Vm.prop.a}) that
\be   \label{eq:Wdet.bd}      
W_{\det}  (x) \geq \frac{1}{\sqrt{2}} V_{2N - 3} \Big(
\frac{x}{\sqrt{2}} \Big)   
\ee

To verify (\ref{eq:V.Slat}) recall that the   
$\gamma_j $'s are normalized and mutually orthogonal.  
Thus
\be       
\lefteqn{\frac{1}{N!} \langle \gamma_0 \wedge \cdots \wedge 
  \gamma_{N - 1} ,   \frac {1} {|{\bf r}_j |} \,  
\gamma_0 \wedge \cdots \wedge \gamma_{N - 1} \rangle } 
 ~~~~~~~~~~~~~~~~~~\nn \\
      & = & \frac{1}{N} \sum_{k = 0}^{N - 1}   
\langle \gamma_k , \frac {1}{|{\bf r}_j |} \,  
\gamma_k \rangle   \\     \nn    
& = & \frac {1}{N} \sum_{k = 0}^{N - 1} V_k (x) = V_{\av} (x)   < 2 V_N(x)
\ee  
where we used (\ref{eq:Vm.prop.d}) in the last line.
The expression (\ref{eq:Wdet}) for $ W_{\det} (x) $ is proved
in the Appendix.

\begin{cor} \label{cor:slat}  Let $h^{\det}(N,Z,B^{-1/2}) $ be the
Hamiltonian described above.  Then for any
$\alpha > 0 $ and any $B $ satisfying   
$a_1 e^{Z^{\alpha/4 }}  > B  \geq a_2 Z^{3 + \varepsilon} $ for 
suitable constants $a_1$ $a_2$ and some $\varepsilon > 0$, there   
exists a constant $A_{\alpha} > 0 $    
\begin{equation}  \label{eq:Slat.Nmax.a} 
N_{\max} (Z, B ) \leq   4Z + A_{\alpha} Z^{1 + \alpha} .
\end{equation}   
Moreover, when $B = a Z^p$ for some $p > 3$,  one can find a constant
$A$ (depending on $a$, $p$) such that
\begin{equation}   
N_{\max} (Z, B ) \leq   6Z + A Z ( \log Z)^2 .
\end{equation}   
\end{cor}


\subsection{Other models} 

For any fixed choice of Landau functions
$\gamma_{m_1} \ldots \gamma_{m_N}$ the effective 
potentials $\wtd{V}^{\up}$ and $\wtd{W}^{\up}$
can be computed  explicitly for both the
case of a simple product and that of a Slater
determinant.   However, general formulas\footnote{In 
the special case, 
$\Upsilon = \prod_{k=1}^N  \gamma_{m_k}^B(y_k,z_k) $ with all 
$m_k$ odd,  $\wtd{W}^{\up}$
is given by a convex sum  which contains
only $V_{2j}$ with {\em even} subscripts as in (\ref{eq:conv.even}).}
   are not so easily obtained.    
Nevertheless, the results in
the appendix hold rather generally in the sense
that $\wtd{W}^{\up}$ is a convex combination
of the form 
$\sum_{i = 0}^{J} b_i  \1rt2 V_{i} \big(\frac{x}{\sqrt{2}} \big) $
with $J \approx 2 \max_k |m_k|$.   

Thus, one might hope to obtain bounds on the effective
potentials similar to those in (\ref{eq:eff.bd})
 but without the constraint $  \nu \leq 2 \mu$.
In fact, one easily finds 
 $\wtd{W}^{\Upsilon}(x) \geq  \1rt2 V_{\nu-1}
\Big(\frac{x}{\sqrt{2}} \Big)$ with $\nu = 2 \max |m_k|$.
However, bounds better than  
$\wtd{V}^{\Upsilon}(x) \leq  V_1(x)$ are not so easily obtained.
In situations in which bounds of the form
\be
  \wtd{V}^{\Upsilon}(x) \leq  c V_{\mu(N)}(x) , \hskip1cm
  \wtd{W}^{\Upsilon}(x) \geq  \1rt2 V_{\nu(N)-1}
   \Big(\frac{x}{\sqrt{2}} \Big) 
\ee
hold with the dependence of $\mu$ and $\nu$ on $N$ known,
this would lead to similar bounds on $N_{\max}(Z,B)$ 
with the contributon of $2Z$  to the linear term replaced 
by one of the form   $\kappa Z$ with $\kappa$
depending on the relative size of $\mu(N)$ and $\nu(N)$.

\section{Proofs} 

\subsection{Localization} \label{sect:local}

The proof of the main theorem will use the RS
localization method which is summarized in \cite{CFKS}.
The argument used here requires some refinements
discussed in more detail  in   \cite{Rusk1, Rusk2}.
   
Let $G_0, G_1, \cdots , G_N $ denote a partition of unity 
consisting of functions 
which are Lipschitz continuous on ${\bf R}^N $
and satisfy $\sum_{j = 0}^N G_{\nu}^2(x) = 1$
as well as the following additional properties:
\begin{enumerate}
    \renewcommand{\labelenumi}{\theenumi}
    \renewcommand{\theenumi}{(\roman{enumi})}

 \item   
 $ \supp(G_0) \subset \{ x : || x ||_{\infty} \leq (1 + \delta)\rho \} $,
     
 \item   
 $ \supp(G_{k}) \subset \{ x :  || x ||_{\infty} \geq \rho  , |
  x_{k} | > \frac{1}{1+ \delta}  || x ||_{\infty} \} $ for
  $   1 \leq k \leq N $, 

\item $\sum_{j = 0}^N |\nabla G_{i} |^2 < \lambda 
   \dfrac{ (\log N )^2}{\delta^2 \,\rho^2}$
     ~ on ~ $\supp(G_0)$, and

\item $\sum_{j = 0}^N |\nabla G_{i} |^2 < \lambda  
   \dfrac{(\log N )^2}{\delta^2 \, x_k^2 }
    \leq   \lambda \dfrac{ (\log N )^2}{\delta^2 \, \rho \, |x_k| }$
    ~  on ~$\supp(G_k)$,
\end{enumerate}
where $\lambda$ is a constant and    $\nabla $ 
denotes the gradient in all the variables $x_1, \dots , x_N $.
The existence of a partition with these properties is
guaranteed by the constructions in \cite{Rusk1, Rusk2}. 

In many applications, one wants a ``sharp'' localization
which is achieved by choosing $\delta$ so that
$\delta \raw 0$ as $Z \raw \infty$, e.g, $\delta = Z^{-\alpha}$
for some $\alpha > 0$.  In such situation, it often
suffices if $|| x ||_{\infty} \leq 2 \rho $ on $\supp(G_0)$.
In this paper, the term ``localization'' is a bit of a 
misnomer, as the radius $\rho$ of the ``inner ball''
will grow with $Z$.   In this case, it can be advantageous
to let the localization be far from sharp and even permit
$\delta$ to grow  with $Z$.
   
Now let $h(N, Z, B^{-1/2} ) $ be as in (\ref{eq:1dimHam}), and
note that the IMS localization formula \cite{CFKS} implies
that for any $\Phi (x_1, \cdots , x_N ) $ in the domain of 
$h = h(N, Z, B^{-1/2})$,    
\be    
\langle \Phi  , h \, \Phi \rangle & = &  
\sum_{\nu = 0}^N \langle  G_{\nu}\Phi  , h \, G_{\nu} \Phi
\rangle - \bra \Phi, LE(x) \, \Phi \ket \\    & = &
\sum_{\nu = 0}^N \langle G_{\nu} \Phi   , 
   \big[ h(N, Z, B^{-1/2})  - LE(x) \big] \, G_{\nu} \Phi \rangle ,   
\ee  
where $LE(x) $ denotes the localization error    
\be   
LE(x) = \sqrt{B} \, \sum_{\nu = 0}^N |\nabla G_{\nu} (x) |^2.   
\ee 
It follows from properties (iii) and (iv) of $G_k$ that 
the localization error is bounded above by
\bsq \be
  L_0 & \equiv & \lambda \frac{ \sqrt{B} \, (\log N )^2}{\delta^2
\,\rho^2} \quad  \hbox{ on } ~   \supp(G_0) \\
   L_k & \equiv & \lambda \frac{ \sqrt{B} \, (\log N )^2}{\delta^2 ~ \rho
\, |x_k| }  \quad  \hbox{ on } ~ \supp(G_k),   \quad k = 1 \ldots N.
   \label{eq:Lk}
\ee  \esq  
To prove Theorem~\ref{thm:main} it suffices to
show that
\be  \label{eq:pregoal}
 \bra  G_k \Phi  \big[ h(N, Z, B^{-1/2} ) - L_k \big] G_k \Phi \ket
   \geq  e_0(N-1,Z,B) \| G_k \Phi^2 \|
\ee
for $k = 0, 1 \ldots N$ with   $e_0(N,Z,B) $  the ground state energy
of $h(N, Z, B^{-1/2} )$.  Since (\ref{eq:pregoal}) is  equivalent to
\be  \label{eq:goal}
 \bra  G_k \Phi  \big[ h(N, Z, B^{-1/2} ) - L_k  - e_0(N-1,Z,B) \big] 
  G_k \Phi \ket  \geq  0, 
\ee
it suffices to show that the quantity in square brackets in
(\ref{eq:goal}) is positive on $\supp(G_k)$ for each
$k = 0, 1 \ldots N$.
It is useful to handle the cases $G_0$ (inner ball estimates) and
$G_1 \ldots G_N$  (outer estimates) separately.   
\medskip 

In the next two sections, we use the convention that $c$ and
$C$ denote   constants in the sense that some
constant exists for which the indicated bound holds.  

\bigskip

 
\subsection{Inner ball estimates:}  \label{sect:in}

On the inner ball (i.e., on supp$(G_0)$)  the 1-dimensional Hamiltonian
(\ref{eq:1dimHam}) with effective potentials satisfying
(\ref{eq:eff.bd}) can be bounded by
\be
h(N, Z, B^{-1/2} ) & \geq &
    N \, e_0(1,Z,\sqrt{B}) + \tfrac{N(N-1)}{2 \sqrt{2}} \,V_{ \nu \mm 1}
  \big( \tfrac{2 \, (1 \! + \! \delta) \,\rho}{\sqrt{2}} \big) \\
  &  \geq & 
- C N \frac{Z^2}{\sqrt{B}}
    \big( \log \tfrac{Z^2 }{B } \big)^2 + \tfrac{N(N-1)}{2}
       \frac{1}{\sqrt{4 (1 + \delta)^2 \rho^2 +   2 \nu} }  \label{eq:in2}
\ee
where we used 
 $h(N, Z, M ) \geq N e_0(1,Z,M)  + \sum_{j < k} \wtd{W}_{jk}^{\up}
    (|x_j - x_k|) $, ~ $|x_j - x_k| < 2 (1 + \delta) \rho$  
on supp$(G_0)$, property (\ref{eq:Vm.prop.c}) and
\begin{equation} \label{eq:AHS}      
- \sqrt{B} \frac{d^2}{dx^2} -   Z V_{\mu} (x) \geq 
   e_0(1,Z,\sqrt{B}) \geq - C
  \frac{ Z^2}{\sqrt{B}} \big( \log \tfrac{ Z^2}{B} \big)^2.   
\end{equation}   
The lower bound in (\ref{eq:AHS}) above
follows from the asymptotic formula 
in \cite{AHS} for the ground state energy  of
the one-electron hamiltonian on the left in (\ref{eq:AHS}).
Now (ignoring the difference between $N$ and $N \mm 1$), 
the right side of
(\ref{eq:in2}) will be positive if 
\be  \label{eq:BZ3cond}
  (1 + \delta)^2  \rho^2 + \frac{\nu}{2} < \frac{1}{C }
  \frac{N^2 B}{Z^4 \big( \log \tfrac{Z^2 }{B } \big)^4} .  
\ee
Since we can assume $Z \leq N$, the right side of
(\ref{eq:BZ3cond}) will be greater than $\nu$ for sufficiently large 
$N,Z$ if
\be   \label{eq:nu.dep}
  B > \frac{\nu}{N} Z^{3 + \epsilon}
\ee
for some $\epsilon > 0$.
For bosonic models  we are be primarily interested in the
case $\nu = O(1)$ for which (\ref{eq:nu.dep}) holds if
$B > Z^{2 + \epsilon}$;  for anti-symmetric models, 
$\nu = O(N)$ so that we need $B > Z^{3 + \epsilon}$.
For now, we assume that  (\ref{eq:nu.dep}) holds, in which case the
requirement (\ref{eq:BZ3cond})  can be rewritten as
\be  \label{eq:RHS}
  (1 + \delta)^2 \rho^2  < \frac{1}{ C }
  \frac{N^2 B}{Z^4 \big( \log \tfrac{Z^2 }{B } \big)^4} - \frac{\nu}{2}
~  <  \frac{1}{2 C }
  \frac{N^2 B}{Z^4 \big( \log \tfrac{Z^2 }{B } \big)^4} .  
\ee
Thus we can ensure that the right side of
(\ref{eq:in2}) is positive by choosing
\be   \label{eq:rho}
    \rho = c \, \frac{1}{1 + \delta} \, 
 \frac{N \sqrt{B}}{Z^2 \big( \log \tfrac{Z^2 }{B } \big)^2}
\ee
for some constant $c$ independent of $N,Z, B, \delta$.  

 Then, since
$- e_0(N-1,Z,B) \geq 0$, the condition (\ref{eq:goal}) will
hold for $G_0$ if
\be   \label{eq:inn.est}
 - C  N \frac{Z^2}{\sqrt{B}}
    \big( \log \tfrac{Z^2 }{B } \big)^2 + 
       \frac{N^2}{2 \sqrt{ 4(1 + \delta)^2 \rho^2 +   2 \nu}}   - 
   \lambda \frac{\sqrt{B}(\log N)^2}{\delta^2 \rho^2}  > 0.
\ee
The first two terms in (\ref{eq:inn.est}) behave like
$+ C N \frac{Z^2}{\sqrt{B}}  \big( \log \tfrac{Z^2 }{B } \big)^2$
and the third like
$ -  \lambda \big(\tfrac{\delta + 1 }{\delta} \big)^{\!2} \,
  \frac{ (\log N)^2 Z^4 \big( \log \tfrac{Z^2 }{B } \big)^4}
   {\sqrt{B} N^2}$.
Comparing these expressions, we find that (using the assumption
that $N > Z$) control
of the localization error requires
\be  \label{eq:inn.loc}
  \lambda \Big(\frac{\delta + 1 }{\delta} \Big)^{\!2}  ~ < ~  
   \frac{ N }{(\log N)^2 \big( \log \tfrac{Z^2 }{B } \big)^2} 
  ~ \Big( \frac{N}{Z} \Big)^2 
\ee
We  consider two cases of small and large $\delta$
separately.
\begin{itemize}

\item[a)]  When $\delta = Z^{-\alpha}$, the left side of
(\ref{eq:inn.loc}) behaves like $\frac{\lambda}{\delta^2}$.
In this  case (\ref{eq:inn.loc}) holds with $\rho$
given by (\ref{eq:rho}) provided that $N > Z$ is sufficiently
large and  
$B \leq C  e^{Z^{1/2  - \epsilon}}$ for some $\epsilon > \alpha$.

\item[b)]  When $\delta > 1$, we can use the fact that
$\frac{\delta + 1 }{\delta} \leq 2$ to see that (\ref{eq:inn.loc})
holds for {\em any} $\delta > 1$ if 
  $N > Z$ and $B \leq C  e^{Z^{1/2  - \epsilon}}$ for some 
$\epsilon > 0$.   Alternatively, we can eliminate the
upper bound on $B$ by letting $N$ grow with $B$ as well as $Z$.
In particular,   (\ref{eq:inn.loc})
holds for  any  $\delta > 1$ if 
  $N > Z   \log \tfrac{Z^2 }{B } $.

\end{itemize}


   
\subsection{Outer ball estimates:}  \label{sect:out}

For any  $k$ such that $ \ 1 \leq k \leq N $, we can
write   
\be
   h^{\Upsilon}(N,Z, M) = h_k^{\Upsilon}(N \mm 1,Z, M) - \frac{1}{M}
\frac{d^2}{dx_k^2}
  - Z \wtd{V}_k^{\up} (x_k ) + \sum_{j: j \neq k}   
   \wtd{W}_{jk}^{\up} (x_j - x_k )
\ee
with the understanding that $h_k^{\Upsilon}(N \mm 1,Z, M) $ is the 
Hamiltonian  obtained by omitting terms in (\ref{eq:1dimHam})
involving $x_k$, but with potentials defined by the $N$-particle
state $\Upsilon$.  Let $E_{0}^{\Upsilon,k}(N \mm 1,Z, B)$ 
denote the corresponding ground state energy.
Since 
$h_k^{\Upsilon}(N \mm 1,Z, M) \geq E_{0}^{\Upsilon,k}(N \mm 1,Z, B)$ and 
$- \frac{d^2}{dx^2} \geq 0$, it follows that
\be   \label{eq:big.ineq}
 \lefteqn{ \bra  G_k \Phi , \big[ 
    h^{\Upsilon}(N,Z, B^{-1/2}) - L_k - E_{0}^{\Upsilon,k}(N \mm 1,Z, B)
   \big] G_k  \Phi \ket }  ~~~~~~  \\  
 & \geq &   \bra  G_k \Phi , \Big[
  - Z \wtd{V}_k^{\up} (x_k ) + \sum_{j: j \neq k}  
   \wtd{W}_{jk}^{\up} (|x_j - x_k|)    - L_k
\Big] G_k \Phi \ket.  ~~~   \label{eq:RHSk}
\ee
For simplicity, we henceforth omit indices $j,k$ on
$E_0, \wtd{V}, \wtd{W}$ and assume\footnote{Since the
bounds on  $\wtd{V}$ and $\wtd{W}$ are independent of $j$
this is not a significant restriction.   In the
case of distinguishable particles, one need only compare 
to $E_{0}^{\Upsilon,k}(N \mm 1,Z, B)$ for a
particular $k$.   When indistinguishable particles are associated
with an asymmetric product of Landau states,  the full 
wave function must be a linear combination of
states of the form (\ref{eq:psi.lowland}) with  terms
associated with irreducible representations of $S_n$ to yield
the appropriate total symmetry; this is a more complex situation
than the simple product model considered here.}
that $\Upsilon$  is a
symmetrized or anti-symmetrized  product.

Now for $x \in \mbox{supp} \ G_k $ we have that 
 $|x_j - x_k | \leq |x_j | + |x_k | \leq (2 + \delta ) |x_k | $, so
that   
\be  
  \wtd{W}^{\up}(|x_j - x_k |) & \geq &
\frac{1}{\sqrt{2}}   
V_{\nu-1} \Big( \frac{|x_j - x_k|}{\sqrt{2}} \Big)  
  \geq     \frac{1}{\sqrt{2}}
V_{\nu-1} \Big( \frac{(2 + \delta)|x_k|}{\sqrt{2}} \Big)
\nn \\ & \geq &   
   \frac{1}{\sqrt{(2 +   \delta)^2 x_k^2 +  2 \nu }}  
 = \frac{1}{2} \frac{1}{\sqrt{(1 +  \tfrac{\delta}{2})^2 x_k^2 +
\frac{\nu}{2} }},
\ee
where we used   (\ref{eq:Vm.prop.a}) and (\ref{eq:Vm.prop.c}).

\medskip

It follows from the upper bound in (\ref{eq:Vm.prop.c}) that 
\be
 \wtd{V}^{\up} (x_k )  \leq   V_{\mu}(x_k) \leq
     \frac{1 }{\sqrt{x_k^2 + \mu}}
\ee
Thus
\be   \label{eq:out.effpot}
- Z \wtd{V}^{\up} (x_k ) + \sum_{j: j \neq k}   
   \wtd{W}^{\up} (x_k )   \geq  - \frac{  Z}{\sqrt{x_k^2 + \mu}}
  + \frac{N \mm 1}{2} \frac{1}{\sqrt{(1 +  \tfrac{\delta}{2} )^2 x_k^2 +
\frac{\nu}{2} }} 
\ee
If one ignores $\delta$, 
the right side of (\ref{eq:out.effpot}) is approximately
\be 
   \frac{-   Z}{\sqrt{x_k^2 + \mu}} + \frac{N}{2}  
  \frac{1}{\sqrt{x_k^2 + \frac{\nu}{2} }} 
\ee
which will be positive if  $N > 2Z$ and $\nu < 2 \mu$.

This explains the origin of the linear term in Theorem~\ref{thm:main}.
It remains to take into account the localization error (\ref{eq:Lk}).
When $\delta \raw 0$ one need only choose $N - 2Z$ large enough
to control (\ref{eq:Lk}).  When larger choices of $\delta $ are
made, one pays the
price of an increase of  $Z  \delta$ in the electrostatic 
estimates, as shown in (\ref{eq:Ncond.b}) below.

 Substituting (\ref{eq:out.effpot}) in
(\ref{eq:RHSk}) and using the estimate (\ref{eq:Lk}),
one finds that (\ref{eq:RHSk}) is bounded below by  
\be   \label{eq:goal.k}
 \bra  G_k \Phi ,  \frac{ T(x_k)}
  {\sqrt{(1 +  \tfrac{\delta}{2} )^2x_k^2 + \frac{\nu}{2}}} \, 
  G_k\Phi\ket,
\ee  
where
\be  \label{eq:goal.T}
  T(x) =  -   Z \sqrt{ \frac{(1 +  \tfrac{\delta}{2} )^2 x^2
  +  \frac{\nu}{2}}{x^2 +\mu}}  + \frac{      N - 1 }{2}  
  - \lambda \frac{\sqrt{B} (\log N )^2}{\delta^2 \rho}  
  \sqrt{(1 +  \tfrac{\delta}{2} )^2 +\frac{\nu}{2x^2}}  
\ee
The expression (\ref{eq:goal.k})
 will be positive, ensuring that
(\ref{eq:goal}) holds for $k = 1 \ldots N$, if $T(x) > 0$ for
$|x|  > \rho $. Since $|x_k| \geq \rho$
on supp$(G_k)$, we find that, with $\rho $ given by (\ref{eq:rho}), 
\be
  \frac{\nu}{2  x_k^2} \leq \frac{\nu}{2N} \frac{Z^3}{B} \frac{Z}{N}
       \big( \log \tfrac{Z^2 }{B } \big)^4 \leq 1
\ee
when $N > Z$ and 
(\ref{eq:nu.dep}) holds.  Thus, we can conclude that 
\be \label{eq:Ncond.a} 2 \, T(x) & \geq & N - 1 - 2   Z \sqrt{
\frac{(1 +  \tfrac{\delta}{2} )^2
   x^2 +  \frac{\nu}{2} }{x^2 + \mu}}
  - 2 \lambda \frac{(1 + \delta)^2(\log N)^2 ( \log \tfrac{Z^2 }{B })^2
Z^2}{ \delta^2 N} \nn
\\ & \geq &  N - 1 - 2   Z \big(1 +  \tfrac{\delta}{2} \big)
  - 2 \lambda \frac{(\log N)^2 ( \log \tfrac{Z^2 }{B })^2 Z^2 }{N}
   \frac{(1 +\delta)^2 }{\delta^2 }
  \label{eq:Ncond.b}
\ee where the second inequality used the assumption $\nu < 2 \mu$.

Now, we can analyze the $N$-dependence of the
right side of (\ref{eq:Ncond.b}) by writing it in the form
\be 2 \, T(x) \geq N - \frac{(\log N)^2 Q }{N} - R
\ee
where $Q,R $ are positive and may depend on $Z, B, \delta$
but are independent of $N$.
The expression on the right is increasing in $N$.  Hence,
for any fixed choice  of $Z, B, \delta$,  if
it is positive for some critical $N = N_c$, then it will be positive
for all $N > N_c$.


\subsection{Completion of proofs} \label{sect:finpf}

To prove Theorem~\ref{thm:main},  choose $N_c = 2   Z+ 1 + a Z^{1
+ 2 \alpha}$. Then \be 2 \, T(x) \geq    Z \bigg[a Z^{2 \alpha} -
c
   \frac{ [(1 + 2 \alpha) \log Z]^2 ( 2 \log Z - \log B)^2}
   {\delta^2 \, Z^{2 \alpha}} -  \delta \bigg]
\ee Thus if $\delta = O(Z^{-\alpha})$, then for sufficiently large
$Z$, \be 2 \, T(x) \geq Z \bigg[ a  Z^{2 \alpha}  - c
   (\log Z)^2 ( 2 \log Z - \log B)^2 -  Z^{- \alpha} \bigg]
\ee which can be made positive for sufficiently large $Z$ as long
as $\log B < C \frac{Z^{\alpha}}{\log Z} $, for some suitable
constant $C . $  This will be true if
 $B < a_1 e^{Z^{\alpha / 2 } } $for some constant $a_1$.
 Replacing $\alpha$ by $ \alpha/2$ yields Theorem~\ref{thm:main}.


To improve the $Z^{\alpha} $ growth of $N$ to one involving
only logarithmic terms, as in Theorem~\ref{thm:log}, one can not
let $\delta \raw 0$.  Moreover, the term 
$\frac{(\delta +1)^2}{\delta ^2}$ 
in (\ref{eq:Ncond.b}) implies that one can not decrease
 the localization error by choosing $\delta $ large.
Therefore, we simply take $\delta = 1$, in which case  
(\ref{eq:Ncond.b}) becomes
\be
\label{eq:Ncond2} 2 \, T(x) \geq N - 1 - 2 Z  -  Z  
  -  c \frac{(\log N)^2 ( \log \tfrac{Z^2 }{B })^2 Z^2}{ N}
. \ee
We now set $N = N_c$ in (\ref{eq:Ncond2}) and
write $N_c = 2Z + 1 + Z f(Z,B) $.  Then
\be \label{eq:delta.est}
  2 \, T(x) & \geq & Z   \Big[ f(Z,B) -    1  - c
    \frac{( \log \tfrac{Z^2 }{B })^2 \big(\log Z + \log f(Z,B) \big)^2}
   {f(Z)}  \Big].
\ee The inner ball estimates in section~\ref{sect:in} require $B =
O(\exp{Z^{1/2 - \epsilon }} )$.  Under this assumption,
Theorem~\ref{thm:main} implies that $f(Z,B)$ grows more slowly than
$Z^{\alpha}$ for some suitable $\alpha $, so that  
$\log f(Z,B) \leq \log Z$ for sufficiently large $Z$. Hence we can find a
constant $A$ such that the right side of (\ref{eq:delta.est}) is positive
when
$f(Z,B) = A \big| \log \tfrac{Z^2 }{B } \big| \log Z$. (Note that
$|\log (Z^2 / B ) | $ will stay bounded away from 0, due to the
lower bound $B
> Z^{\gamma _{\nu } } $, $\gamma _{\nu } > 2 $, which
ensures that $f(Z,B) > 1 $ for big $Z . $) This proves
Theorem~\ref{thm:log}

Since our hypotheses do not permit $B$ to grow exponentially with
$Z$ and the case of greatest interest is polynomial growth, e.g.,
$B = Z^{3 + \varepsilon}$, it is useful to restate our results under
the assumption that $B = Z^p$ for some $p >0$.  In that case, we
can conclude that there is a constant $A$, depending on $p $, such
that (\ref{eq:delta.est}) is positive for $f(Z) = A (\log Z)^2 $,
which proves Theorem~\ref{thm:Zp}.  This also gives a bound of
   $N_{\max}(Z,Z^p) \leq 2Z +
     A  Z (\log Z)^2 . $

Corollary~3 follows immediately from Theorem~\ref{thm:log}
and the discussion above. 


To prove Corollary~\ref{cor:0}, it suffices to observe
that the hypotheses of Theorems 
\ref{thm:main} and \ref{thm:log} hold with
$\mu = m$,  and $\nu = 2m$.   Unless $m$ depends upon $N$,
 we now have
$\nu = O(1)$ so that   (\ref{eq:nu.dep}) holds
if $B > Z^{2 + \varepsilon}$.


To prove Corollary~\ref{cor:slat} 
 for the Slater model, note that it follows
from (\ref{eq:Vm.prop.d}) and (\ref{eq:V.Slat} that 
$\wtd{V}^{\up}(x) = V_{\av}(x) \leq 2V_N(x)$,
and from (\ref{eq:Wdet.bd}) that
 $\wtd{W}^{\up}(x) \geq V_{2N-3}(x)$.
Thus, $h^{\det}(N,Z,M)$ satisfies that hypotheses
of Theorems \ref{thm:main} and \ref{thm:log} with  
$\mu = N$, $\nu = 2N-2 $, but with
  $Z$ is {\em replaced by an effective charge} of $2Z$.  
This has the effect of doubling the coefficient 
in the linear term and
modifying the constant in the non-linear term.
Since $\nu$ is $O(N)$, (\ref{eq:nu.dep}) requires
 $B > Z^{3 + \epsilon}$.

\rmk With more technical effort some of the hypotheses
can be relaxed and/or estimates improved as sketched below.

\begin{enumerate}

\item The condition $\nu \leq 2 \mu$ is needed only for
the bound 
 $ \sqrt{ \frac{(1 + \tfrac{\delta}{2})^2x ^2 + \frac{\nu}{2} }{x^2 +\mu}}
\bigg|_{x \approx 0} \leq 1$ implicitly used in (\ref{eq:Ncond.b}). 
However, this  is actually used only in analyzing the outer region for
which one can assume $|x_k| \geq \rho$ is much larger than $0$.  Hence, 
with a bit more effort, this condition can probably be
dispensed with.  At worst, allowing $\nu > 2 \mu$ 
would only change the coefficient of the (lower order)
linear term. 

\item The upper bound $B \leq C  e^{Z^{1/2  - \epsilon}}$  is somewhat
artificial.   It arises because we have chosen to state our results in a
way that emphasizes the dependence of $N_{\max}(Z,B)$ on $Z$.  More
rapid growth of $B$ will increase the confinement of the
electrons in two dimensions, but make them more
delocalized in the direction orthogonal to the field.
Hence, it is not surprising that the localization error
will be harder to control if $B$ grows exponentially with $Z$.

 In the case of
Theorem~\ref{thm:log}, one can eliminate the need for
this  upper bound  in controlling the localization error
on the inner ball by using the fact that 
$\frac{N}{Z} > \log \frac{Z^2}{B}$.
This was  noted in remark (b) after (\ref{eq:inn.loc}).
However, if $B$ grows exponentially with $Z$, then the
estimate $\log f(Z,B) < \log Z$ used after (\ref{eq:delta.est})
will no longer be valid.   The upper bound can be eliminated
by allowing $N_c$ to grow sufficiently with $B$.  The choice
$f(Z,B) = (\log Z)^2 + \log Z (\log B)^{1 + \omega}$ for some
$\omega > 0$ will suffice.  This proves the result
 stated as Theorem 3.1 in \cite{BRW} and given after
  (\ref{eq:gform}) in
the introduction.

\item The linear term in the Slater model is doubled because
we use the estimate  $V_{\av}(x) \leq  2 V_N(X) $
which gives an effective charge of $2Z$ rather than $Z$. 
Although this bound is tight  near $x \approx 0$, it
is used only in the outer ball where $|x_k| > \rho$ and
one would expect  $V_{\av}(x) \approx  V_N(x) \approx V_{N-1}(x) $. 
(Note that since $\nu = 2(N \mm 1)$ it would suffice to have
a bound with $\mu = N \mm 1$.)

In fact, using results in \cite{BRW,RW} one can show that
\be
  V_{\av}(x)   & = & 2V_N(x) - \frac{2x^2}{N} \Big(
   \frac{1}{x} - V_{N-1}(x) \Big)  \\ 
& \approx &   2 V_N(x) - \frac{1}{\sqrt{x^2 +N} }
    + O \Big(\frac{N}{x^2} \Big) .
\ee
When $|x| > \rho$, and $B > Z^3$ this becomes
\be
  V_{\av}(x)  &  \approx  &  2 V_N(x) - \frac{1}{\sqrt{x^2 +N} }
    + O \Big(\frac{ (\log Z)^4}{ N} \Big) \nn \\
    & \leq & V_N(x) + C \frac{ 1}{Z}  
\ee
for sufficiently large $Z$ and $N > Z^{1 + \alpha}$
Thus, one could expect to show that (\ref{eq:Slat.Nmax.a})
can be replaced by $N_{\max} (Z, B ) \leq   2Z + A_{\alpha} Z^{1 +
\alpha}  $ in Corollary~\ref{cor:slat} by using more
refined estimates for $V_{\av}(x)$.  This would give  the
expected behavior of $2Z$ for the linear term when
$B$ grows more rapidly than $Z^3$.  

\end{enumerate}

 \pagebreak

\section{Discussion}  \label{sect:conc}

 The first step of the
RS method is to divide the system  into a small
``inner'' ball in which binding is precluded because the electrons are
confined to a small region, and an ``outer'' ball in which the
localization error becomes negligible as $Z \rightarrow \infty$.  For
bosonic systems, one expects to be able to squeeze the electrons closer
together, yielding a smaller cut-off $\rho$ than for fermions.  This
feature is the {\em only} factor which precludes extending the
proof of asymptotic neutrality in  \cite{LSST} to bosonic atoms.
This suggests that the localization error is not simply a
technical artifact, but a reflection of a real physical effect.
In the one-dimensional models considered here, the anti-symmetry
required by the fermionic
nature of electrons is achieved entirely within the
Landau band.   This results in a in a one-dimensional
model that is  bosonic, with the anti-symmetry 
 reflected only in the effective potentials 

The one-dimensional confinement also delocalizes the electrons.
  This is reflected by the  the effective mass of
$M = B^{-1/2}$ in (\ref{eq:1dimHam}) which implies that  
 in strong fields the electrons behave
like extremely light  particles.  The uncertainty principle
then implies that trial wave functions which localize the
electrons cannot yield bound states.   Thus, it may seem
rather surprising that localization methods can be applied
successfully.  For atoms in strong magnetic fields, 
this terminology may be misleading because the cut-off
radius  $\rho$ is not small.  Instead 
$\rho \sim N \sqrt{B} Z^{-2} (\log \frac{Z^2}{B})^{-2}$ which
grows with $B$.  Thus,  localization methods can be used to obtain  
(non-optimal) upper bounds on $N_{\max}$ despite the fact that the
electrons are highly delocalized and the size of the ``inner'' region
becomes  large as $B \rightarrow \infty$.  

Lieb's method can be interpreted as
 a different type of localization in which
$G_k$ is essentially the inverse square root of the potential.
In three dimensions, the resulting localization error can be
completely controlled by the kinetic energy, eliminating the
need for an additional inner/outer deloclization provided that
the magnetic field goes to zero at infinity.
However, as discussed in section 3 of \cite{BR1}, control of
the localization error is more complex in models resulting
from the types of magnetic fields considered here.

When Lieb's method was applied to a one-dimensional
model   in \cite{BR1} control of the localization error
led to a $\sqrt{B}$ growth in $N$.  In \cite{S} Seiringer
showed that better bounds can be obtained if one  
applies Lieb's method to the full 3-dimensional
Hamiltonian, yielding results comparable to those
obtained here. 

One might try to combine Lieb's method with the inner/outer
localization used here, i.e., in Section~\ref{sect:local}
use $G_0$ as in (i) but in (ii) replace $G_k$  by 
$\sqrt{ \frac{1 - G_0^2}{\wtd{V}^{\up}(x_k)} }$ for $k \geq 1$. 
In the case of the simple $0$-model, the argument given  in Section 4 of
\cite{BR1}  can be used in the outer region with the term
$\ds{ \lim_{x \raw 0}} \tfrac{|\nu^{\prime}(x)|^2}{4 \nu(x)}$
replaced by $\frac{|\nu^{\prime}(\rho)|^2}{4 \nu(\rho)}$. 
With $\rho$ as in Section~\ref{sect:in}, this would yield a net bound of
\be
 N_{\max}(Z,B) \leq 2Z + A \,Z \,  |\log \tfrac{Z^2}{B} |
\ee
which is a very slight improvement.\footnote{
The change  from a quadratic to a linear dependence on $\log Z$
is due to the fact that the LE arising from $G_0$ does not require
 a $(\log N)^2$ in the numerator in (iii) and (iv), resulting in a net
bound of the form $\frac{\lambda}{\rho^2}$.}  To
extend this to $m \neq 0$, would require additional
work.  However, for $|x| > \rho$ one should be able to show
that $[V_m(x)]^{-1} \approx \sqrt{x^2 + m}$ to obtain 
a similar bound.   In the case of the Slater model, one would
also need estimates of the type discussed in Remark~3.

\bigskip


\noindent{\bf Acknowledgment:}  This work was started when the
RB was professor at the University of Reims, France, and MBR
a visiting professor there.  Some of this
work was done while both authors were visiting the Schr\"odinger 
Institute at the University of Vienna.  The final version was written
while MBR was a Walton visitor at the Communications Network Research
Institute of Dublin Institute of Technology.  It is a pleasure to
acknowledge the hospitable   environment at these institutions.

\bigskip

\appendix

\section{Appendix} \label{apx}

We begin by proving a special case of   Pr\"oschel,
et al's  result  \cite{PRWRH}  that the effective
interaction $W_{mm\'}(x_1 -x_2)$ can be written as a convex
combination of potentials of the form 
$\1rt2 V_j \big( \tfrac{x_1 -x_2}{\sqrt{2}} \big) $
with $j \leq m + m\'$.  In the special case $m = m\'$, only terms with
even subscript occur in the convex combination.
\begin{lemma} \label{lemm:Wm}
The effective interaction $W_{mm}(x_1 -x_2)$ defined in 
(\ref{eq:conv.mm}) satisifies
\be   \label{eq:conv.even}
  W_{mm}(x_1 -x_2) = \sum_{j = 0}^{m} 
   b_{2j} \1rt2  V_{2j} \big( \tfrac{(x_1 -x_2)}{\sqrt{2}} \big) .
\ee
with $b_{2j} > 0$ and $\sum_{j = 0}^{m} b_{2j} = 1$. 
\end{lemma}
\pf Substituting for $\gamma_m$ in (\ref{eq:Weff}) and
writing out the resulting integral yields
\be  \label{eq:Wmm}
 W_{mm}(x_1 -x_2) = \frac{1}{\pi m!}
   \int_{\bf C} \int_{\bf C} d\zeta_1  d\zeta_2
  \frac{ |\zeta_1|^{2m} |\zeta_2|^{2m} e^{-|\zeta_1|^2} e^{-|\zeta_2|^2}}
{\sqrt{(x_1 -x_2)^2 + |\zeta_1 -\zeta_2|^2}}.
\ee
We now make the complex change of variables to 
\be  \label{eq:chg.var}
 \sigma \equiv \1rt2 (\zeta_1 + \zeta_2)   & ~~ & 
  \tau \equiv \1rt2 (\zeta_1 - \zeta_2) \\
s = |\sigma|= \1rt2 |   \zeta_1 + \zeta_2| & ~~ & 
  t = |\tau| = \1rt2 |\zeta_1 - \zeta_2|  \nn
\ee
 and let  $\theta$ 
be the angle between $\sigma$ and $\tau$.  Then
$|\zeta_1 |^2 + |\zeta_2 |^2 = s^2 + t^2 $ and
\be
  |\zeta_1|^2 |\zeta_2|^2 & = &
   (s^2 + t^2 + 2st \cos \theta)  (s^2 + t^2 - 2st \cos \theta) \nn \\
  & = &  s^4 + t^4 - s^2 t^2 \cos  2 \theta .
\ee
Substituting in (\ref{eq:Wmm}) yields
\be  \label{eq:75}
 W_{mm}(x_1 -x_2) = \frac{2}{m!}
   \int_0^{\infty} \! e^{-t^2} t dt \int_0^{\infty} \! e^{-s^2} s ds
    \int_0^{2 \pi} \! \!  d\theta \,
  \frac{ \big( s^4 + t^4 - s^2 t^2 \cos  2 \theta \big)^m}
   {\sqrt{(x_1 -x_2)^2 + 2 t^2}}.
\ee 
Performing the integral over $s$ and $\theta$ yields
\be  \label{eq:Wmm.poly}
 W_{mm}(x_1 -x_2) = \int_0^{\infty} \! e^{-t^2} t dt
  \frac{P(t^2)}
   {\sqrt{(x_1 -x_2)^2 + 2 t^2}}
\ee
for some polynomial $P(u)$ of degree $2m$.  Writing
$P(u) = \sum_{i = 0}^{2m} b_i u^i $, and substituting in
(\ref{eq:Wmm.poly}) immediately yields
\be
  W_{mm}(x_1 -x_2) = \sum_{i = 0}^{2m} b_i \1rt2 
   V_i \big( \tfrac{ x_1 -x_2 }{\sqrt{2}} \big) .
\ee
It remains only to show that the coefficients $ b_i$ are
even, positive and sum to one.  Applying the binomial
expansion to the numerator in (\ref{eq:75}), one
easily sees that terms with $i = 2j$  have positive
coefficients $b_{2j} > 0$.  When $i = 2j +1$ is odd,
one has an integral of the form 
$\int_0^{2 \pi} \cos^{2j+1} 2 \theta \,  d\theta = 0 $
which implies $b_{2j+1} = 0$.  Thus, only the 
coefficients with $i = 2j$ survive, and these
are positive.
To see that $\sum_{j = 0}^{m} b_{2j} = 1$, it suffices
to use the fact that both $W_{mm} (x ) $ and all $V_i (x ) $ behave
like $1 / |x | $ at infinity.

When $m \neq m\' $ the integrand will  contain
an additional factor of the form 
$(s^2 + t^2 + 2st \cos \theta)^{|m - m\' |}$.
As above, one can show that terms involving
$\cos^j \theta$ integrate to zero when $j$ is odd.

One can use a similar strategy
to show that any inner product of the form
\bsq \be  \label {int:direct}
& \bra \gamma_j(\zeta_1) \gamma_k(\zeta_2 ), \, \frac {1}
{{|\bf r}_1 - {\bf r}_2 |}   \,
 \gamma_j(\zeta_1) \gamma_k(\zeta_2 ) \ket & \hskip1cm \hbox{or} \\
& \bra \gamma_j(\zeta_1) \gamma_k(\zeta_2 ), \, \frac {1}
{{|\bf r}_1 - {\bf r}_2 |}   \,
 \gamma_k(\zeta_1) \gamma_j(\zeta_2 ) \ket &   \label{int:exch}
\ee  \esq
can be written as a linear combination  
$\ds{\sum_{j = 0}^{j+k}} c_j \1rt2  V_{j} 
   \big( \tfrac{(x_1 -x_2)}{\sqrt{2}} \big)$
 with  $\sum_j c_j = 1$.   In the case of
 (\ref{int:direct}), as sketched above and shown in \cite{PRWRH}, 
the coefficients $c_j$ are positive  giving a convex combination.
For the exchange integrals (\ref{int:exch}), this need not be true.
However, for antisymmetric products, the two types of integrals arise
together in combinations whose net coefficients  
  are positive, as shown below.

\begin{lemma}  \label{lemm:det}
For any choice of $0 < m_1 <  m_2  \ldots < m_N$, 
\be   \label{eq:Wmdet}
W_{\det}^{m_1 \ldots m_N} (x_1 - x_2 ) & = &
\tfrac{1}{N!}
  \langle \gamma_{m_1} \wedge \cdots \wedge \gamma_{m_N} ,
   \frac {1}{|{\bf r}_1 - {\bf r}_2 |} \,
\gamma_{m_1}   \wedge   \cdots    \wedge \gamma_{m_N} \rangle \nn \\
& = & \1rt2
      \sum_{j = 0}^{J} b_{2j+1} V_{2j+1}\Big(
     \tfrac{x_1 - x_2 }{\sqrt{2}} \Big) .
\ee
with  $ b_{2j+1} \geq 0 $ and $\sum_j b_{2j+1} =1$, and  \
   $J =  m_{N- 1}+ m_N$, 
\end{lemma}

In the Slater model, we have $m_k = k \mm 1$, from which
it follows that $J = 2N \mm 3$.
\begin{cor}  \label{lemm:W}
For the Slater model, the effective interaction in
$h^{\det}(N,Z,B^{-1/2}) $ (\ref{eq:1dimHam})  is  a convex
combination of $V_m$ with odd $m = 1, 3, \ldots  2N \mm 3$, i.e.,
\be \label{eq:VintS}
\wtd{W}^{\up}(x_1 - x_2 ) = W_{\det}(x_1 - x_2 ) = \1rt2
      \sum_{j = 0}^{N - 2} b_{2j+1} V_{2j + 1}\Big( 
  \tfrac{ x_1 - x_2 }{\sqrt{2}} \Big) .
\ee
with    $ b_{2j+1} > 0 $ and $\ds{\sum_{j = 0}^{N \mm 2} b_{2j+1} = 1}$.
\end{cor}


\pf To prove Lemma~\ref{lemm:det}, we first consider the special
case $N= 2$.  Let $j, k$ be fixed, and write
\be  \label{eq:Wint}
W_{\det}^{j,k}(x_1 - x_2 ) & = & 
  2 \langle \gamma_j \! \wedge \! \gamma_k (\zeta_1, \zeta_2 ), \,
\frac {1} {{|\bf r}_1 - {\bf r}_2 |}   \, \gamma_j \! \wedge \! \gamma_k
(\zeta_1 , \zeta_2 ) \, \rangle    \nn
  \\ & = &  \int_{\bf C}  \int_{\bf C}
\frac{|\zeta_1^j \zeta_2^k - \zeta_1^k \zeta_2^j|^2  ~ e^{-|\zeta_1 |^2 - |\zeta_2 |^2}}
   {\sqrt{ (x_1 - x_2 )^2 + |\zeta_1 - \zeta_2 |^2}} d\zeta_1 d\zeta_2
,   
\ee 
where  we used (\ref{eq:landau}) and as before, ${\bf r}_j =
(x_j , y_j ,z_j ) $ and $\zeta_j = y_j + i z_j $. 
Now make a change of variables as in
(\ref{eq:chg.var}).   Using the binomial
expansion,  we find
$$
\zeta_1 ^j \zeta_2 ^k = 2^{-(j + k ) / 2 } \sum _{\nu = 0 } ^j \sum _{\mu
= 0 } ^k (-1 )^{\mu } \tbinom{j}{\nu} \tbinom{k}{\mu} \tau^{\mu +
\nu }
\sigma^{j + k - (\mu + \nu ) } ,
$$
so that \be \label{eq:w1w2} \zeta_1 ^j \zeta_2 ^k - \zeta_1 ^k \zeta_2 ^j = \sum
_{\alpha = 0 } ^{j + k } A_{\alpha } \tau^{\alpha } \sigma^{j + k
- \alpha } , \ee 
where 
$$
A_{\alpha } = 2^{- (j + k )/2 } \sum _{\begin{array}{cc} \nu + \mu =
\alpha \\
\nu \leq j, \mu \leq k \end{array} } \Big[ (-1 ) ^{\mu } - (-1
)^{\nu } \Big]) \tbinom{j}{\nu} \tbinom{k}{\mu}  .
$$
and we suppress  the dependence of $A_{\alpha}$ on $j,k$. 
When  $\alpha $ is even, $(-1 ) ^{\mu } = (-1 ) ^{\nu }$
so that $A_{\alpha } = 0 $.  Therefore, only terms with
$\alpha $ odd survive in the sum  (\ref{eq:w1w2}).  Moreover,
\be  \label{eq:intw1w2}
 W_{\det}^{j,k}(x_1 \mm x_2 ) & = &  \int_{\bf C}  \int_{\bf C}
\frac{|\zeta_1^j \zeta_2^k -
\zeta_1^k \zeta_2^j|^2  ~ e^{-|\zeta_1 |^2 - |\zeta_2 |^2}}
   {\sqrt{ (x_1 - x_2 )^2 + |\zeta_1 - \zeta_2 |^2}} d\zeta_1 d\zeta_2  
     \\
&=& \sum _{\alpha ~{\rm odd} } \sum _{\beta ~{\rm odd} } A_{\alpha } 
  \overline{A}_{\beta}   \int _{\bf C } \int _{\bf C }
\frac{\tau^{\alpha } \overline{\tau}^{\beta } \sigma^{j + k - \alpha }
\overline{\sigma}^{j + k - \beta } }{\sqrt{ (x_1 - x_2 )^2 + 2t^2}} 
 ~ e^{-s^2 - t^2 } \, d\sigma d\tau \nonumber
\ee
Next, write $\tau = t e^{i \varphi}$ and use the fact that
\bee
   \int _{\bf C }
  \tau^{\alpha } \overline{\tau}^{\beta }  f(t) \, d\tau =
  \int_0^{\infty} t^{\alpha + \beta} f(t) dt \int_0^{2 \pi}
   e^{i \varphi (\alpha - \beta)} d\varphi
\eee
is zero if $\alpha \neq \beta$ to see that the integral  after
(\ref{eq:intw1w2}) becomes
\be
(2 \pi )^2 \sum _{\alpha ~{\rm odd}} | A_{\alpha } |^2 \int _0
^{\infty } s^{2(j + k - \alpha) + 1 } e^{-s^2 } ds \int _0
^{\infty } \frac{|t |^{2 \alpha + 1 } \quad e^{-t^2 }  }{\sqrt{(x_1 - x_2
)^2 + 2t ^2 } } ~ dt .
\ee
Integrating over $s$ then yields 
$$
W_{\det}^{j,k}(x_1 - x_2 ) =  \sum _{\alpha~ {\rm odd}  }  
  b_{\alpha} \1rt2   V_{\alpha } \Big( \frac{x_1 -
x_2 }{\sqrt{2 } } \Big) ,
$$
for suitable constants $b_{\alpha } $, of which we only want to
note that they are strictly positive when $\alpha$
is odd and in the range $1 \leq \alpha \leq j + k$.   As before,
 $\sum_j b_j = 1 $ follows
easily from the fact that both $W_{\det} (x ) $ and the $V_j (x ) $ behave
like $1 / |x | $ at infinity. This proves
Lemma \ref{lemm:det} in the case $N = 2$.

The general case then follows from the fact that an $N$-particle
Slater determinant is a convex combination of two-particle
slater determinants.  In fact,
\be    
W_{\det}^{m_1 \ldots m_N} (x_1 - x_2 ) & = &
\tfrac{1}{N!}
  \langle \gamma_{m_1} \wedge \cdots \wedge \gamma_{m_N} ,
   \frac {1}{|{\bf r}_1 - {\bf r}_2 |} \,
\gamma_{m_1}   \wedge   \cdots    \wedge \gamma_{m_N} \rangle \nn
\\    & = & \tfrac{2}{N (N - 1 )} \sum_{j < k}
\langle \gamma_{m_j} \! \wedge \! \gamma_{m_k} (\zeta_1, \zeta_2 ), \, \frac {1}
{{|\bf r}_1 - {\bf r}_2 |}   \, \gamma_{m_j} \! \wedge \! \gamma_{m_k}
(\zeta_1 , \zeta_2 ) \, \rangle.  ~~~ \nn \\
 & = & \tfrac{2}{N (N - 1 )} \sum_{j < k}  W_{\det}^{m_j,m_k}(x_1 - x_2 ).
\ee

\pagebreak

\end{document}